# Binary Decision Process in Pre-Evacuation Behavior

Peng N. Wang Peter B. Luh Peter Sincak Laura Pitukova

Xuesong Lu W. M. Wonham Timo Korhonen

*Abstract*— **In crowd evacuation the time interval before decisive movement towards a safe place is defined as the pre-evacuation phase, and it has crucial impact on the total time required for safe egress. This process mainly refers to situation awareness and response to an external stressors, e.g., fire alarms. Due to the complexity of human cognitive process, simulation is used to study this important time interval. In this paper a binary decision process is formulated to simulate pre-evacuation time of many evacuees in a given social context. The model combines the classic opinion dynamics (the French-DeGroot model) with binary phase transition to describe how group pre-evacuation time emerges from individual interaction. The model parameters are quantitatively meaningful to human factors research within socio-psychological background, e.g., whether an individual is stubborn or open-minded, or what kind of the social topology exists among the individuals and how it matters in aggregating individuals into social groups. The modeling framework also describes collective motion of many evacuee agents in a planar space, and the resulting multi-agent system is partly similar to the Vicsek flocking model, and it is meaningful to explore complex social behavior during phase transition of a non-equilibrium process.**

## I. Introduction

Evacuation behavior is an important human factor in safety performance engineering, and pre-evacuation time is identified as an interval between receiving an alarm signal and decisive escaping for safety. This mental process mainly refers to situation awareness and response to an external alarm signal, and it often makes up a significant portion of the overall evacuation time and thus is crucial to egress efficiency and safety. Existing models and theories to explain crowd pre-evacuation behavior are mainly from the perspective of behavioral science and psychology [1], [2], [3], involving cognitive appraisals as to if the situation is threatening or not before deciding to escape. One of the most well-known theory is called fight-or-flight response, i.e., people choose either fighting or fleeing in response to a threatening situation. In this process leaving for safety is one option, but seldom the first option [1].

Very importantly, if many individuals are involved in evacuation, leaving for safety is rarely an individualistic action, but a group decision in a collective sense. People are inclined to be socially connected as a group to respond to the outside, and such social group behavior helps relieve stress, providing group members with a sense of safety. However, the trade-off is that it may take certain time for individuals to form groups. Opinions of the group members may also vary to some extent such that it also takes time for them to reach a consensus. The group size and interpersonal relationship of its members are among the key factors that determine group dynamics [2], [4], [5]. Things often become better if an effective leader emerges in the group to avoid responsibility diffused or inhibited.

Given this research background, this paper introduces a decision process to describe many individuals' collective response in pre-evacuation phase, i.e., choosing one phase of either defending (fight) or escaping (flight). The model is established as a mixture of physics models and dynamical systems with socio-psychological theories. A novel binary decision process is formulated in Section 2 on the basis of classic opinion dynamics, widely known as the French-DeGroot Model. The entire system is also considered as a set of interacting extended automata for the pre-evacuation dynamics of many individuals. In Section 3 time-varying social topology is discussed and the decision process is further integrated in the Vicsek flocking framework. The concluding remarks are presented in Section 4.

---

*Research was supported in part by National Science Foundation under Grants CMMI-1000495.

Peng N. Wang, Peter B. Luh and Xuesong Lu were with Department of Electrical and Computer Engineering, University of Connecticut, Storrs, USA. (e-mail: wp2204@gmail.com; Peter.Luh@uconn.edu; Xuesong.Lu@uconn.edu).

Peter Sincak and Laura Pitukowa are with Center for Intelligent Technologies, Department of Cybernetics and Artificial Intelligence, Technical University of Kosice, Slovakia, European Union. (e-mail: peter.sincak@tuke.sk; laura.pitukova@student.tuke.sk)

W. M. Wonham was with Department of Electrical and Computer Engineering, University of Toronto, Ontario, Canada. (e-mail: wonham@utoronto.edu.ca).

Timo Korhonen was with Fire safety Research Department in VTT Technical Research Center of Finland, Finland. (email: Timo.Korhonen@vtt.fi)

## II. Opinion Dynamics with Binary Phase Transition

The crowd evacuation activities can be divided into two major phases: pre-evacuation phase and movement phase. In the first phase evacuees often try to interpret the alarm signal received, collecting cues to confirm what really occurs (e.g., false alarm) instead of directly moving to an exit or a safe place. During this process an important issue is how people are self-organized into groups because they normally tend to be socially connected to gain a sense of safety. A decision process is thus formulated as follows to describe situation awareness and response in the pre-evacuation phase.

In the following discussion the term "agent" is often used to represent evacuees. We denote matrices with capital letters, e.g., $\boldsymbol{C}=[c_{ij}]_{n\times n}$, and use lower case letters for scalar entries. Let $n$ be the total number of evacuee agents under consideration. All the agent are sequentially indexed by $i$=1, ... n. Given simulation starting at $k$=0, all the evacuee agents receive the alarm signal at $k$=0. The desired pre-evacuation time for $n$ agents are initialized by $x_i(k)>0$ at initial time step $k$=0. In concept, $x_i(k)$ measures the psychological state of agents, referring to perception of time-pressure or subjective feeling of urgency which is caused from an outside environmental stressor, e.g., an alarm signal. Such perception is individualized for different people may perceive the same situation differently.

In another perspective since $x_i(k)$ reflects perception of the time-pressure, it also fits in physics measurement of time. In real-world scenarios $x_i(k)$ has unit of seconds, minutes or hours, depending on different types of events. For example, in response of building fire alarm, $x_i(k)$ often last for minutes or seconds while in response of bushfires or floods $x_i(k)$ is in unit of days or hours. Consider $t=k\Delta t$ is the physics time, also in unit of seconds, minutes or hours., and $x_i(k)$ for k=0,1,2… reflects how agent $i$ feels the level of urgency expanding in the simulation timeline. At each time stage k=0,1,2…, the agents' opinions evolve as follows.

$$x_i(k+1) = (1-p_i)x_i(k)+ p_i\Sigma_j c_{ij}x_j(k) = x_i(k) + p_i \Sigma_j c_{ij} [x_j(k)- x_i(k)] \qquad 0\le p_i\le 1 \quad 0\le c_{ij}\le 1 \quad \Sigma_j c_{ij}=1 \tag{1}$$

The above dynamical process describes how an agent's desired pre-evacuation time $x_i(k)$ is updated by iteratively mixing itself with a weighted average of others in social connections. Parameter $p_i$ reflects how agent $i$ keeps balance between its own opinion and others, and it indicates whether the agent is stubborn or open-minded. Parameter $c_{ij}$ is non-zero if agent i has access to acquire agent $j$'s opinion $x_j(k)$. The model is primarily a mixture of individualistic and social behavior, describing how individual-level opinions reach consensus in a given social context.

The matrix $\boldsymbol{C}=[c_{ij}]_{n\times n}$ is a quantitative measure of social connection of $n$ agents, specifying weights that any of the agents puts on the opinions of all the others. These weights are summarized compactly into a matrix $\boldsymbol{C}=[c_{ij}]_{n\times n}$, which is the adjacency matrix of a directed graph as shown in Fig 1(a). For instance, an arrow pointing from Agent 3 to Agent 2 means that Agent 2 is socially influenced by Agent 3 with a normalized weight $c_{23}$=0.3. In practical computing $\Sigma_j c_{ij}=1$ enables us to use a gossip-based asynchronous approach to compute the above pre-evacuation dynamics [6].

An issue to be emphasized is on the diagonal element of $\boldsymbol{C}$. We define $c_{ii}$=0 if there exists any input arc from other agent (j≠i) to agent $i$ in the topology. If not, we let $c_{ii}$ =1 to meet the requirement of $\sum_j c_{ij}=1$. Thus, (1) is applicable to either an agent socially influenced by others or an isolated one. As for an isolated agent who has no input connection from others, it implies that $c_{ij}$ =0 for ∀j≠i, and we have $c_{ii}$=1 such that $x_i(k)$ is not changing with time.

Equation (1) looks a little intricate for it decomposes the traditional French-DeGroot model by two sets of parameters, i.e., $c_{ij}$ and $p_i$ . This mathematical structure is useful to discuss time-variant topology since changing $c_{ij}$ is decoupled from $p_i$ , and we are thus able to discuss changing topology $\boldsymbol{C}=[c_{ij}]_{n\times n}$ independently from the stubbornness parameter $p_i$. The resulting graphical model is time-varying as to be discussed in Section 3.

Take three agents for example in Fig. 1, and the directed arc from Agent 1 to Agent 2 means that opinion of Agent 2 is impacted by Agent 1 with the influential weight of $c_{21}$=0.7. The initial values of $x_i(k)$ at k=0 are 13min, 5min and 21min for agent $i$=1,2,3. The decision balance parameter $p_i$ is given by $p_1$=0.3, $p_2$=0.5, $p_3$=0.25. In addition, there is a dash line drawn by $t=k\Delta t=x$. It divides the entire $t$-$x$ plane into two regions. If the simulation step is Δt=1 minute and x is also in unit of minutes, and the axis of $t$ and $x$ are on the same scale (i.e., the entire $t$-$x$ plane is divided into two regions equally by $t=k\Delta t=x$ with Δt=1). As illustrated, the upper left area of the $t$-$x$ plane is the pre-evacuation phase because simulation time $t$ has not yet reached the desired pre-evacuation time ($t<x$). The lower right area corresponds to the movement phase ($t\ge x$). When the colorful lines meet the dash line, phase transition takes place and the corresponding agent starts to move towards an exit or safe place. Thus, the entire figure demonstrates binary phase transition of each evacuee.

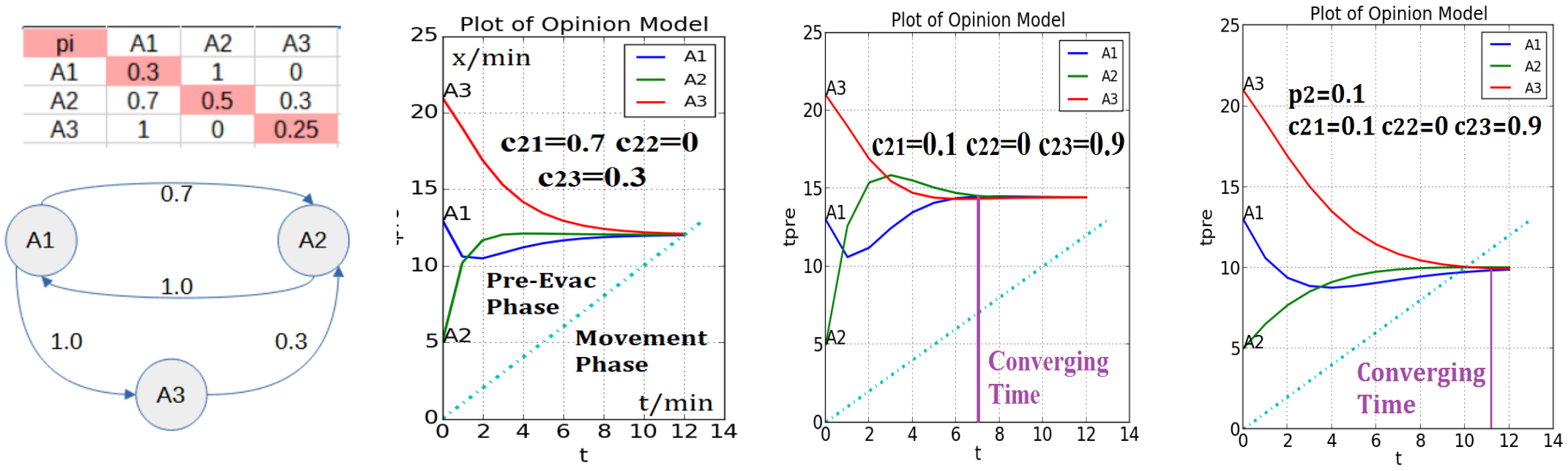

(a) (b) (c) (d)

**Fig. 1. Social Topology of Agents and Opinion Dynamics**

The above model characterizes a kind of binary phase transition between pre-evacuation and movement phases as shown in Fig. 2. It primarily agrees with the fight-or-flight response, and it reflects how time-pressure affects evacuees' decision making in a collective sense. The simulation time $t=k\Delta t$ is a common variable known by all the individual agents. It is uncontrollable in nature because it proceeds irreversibly and irresistibly. The phase transition takes place if $t$ reaches an agent's desired pre-evacuation time i.e., $t \geq x_i(k)$. We can understand this model as a set of interacting extended finite automata with individual variable $x_i(k)$ and a common variable $t=k\Delta t$ shared by all the agents [7], [8]. Agent $i$ is able to adjust its own $x_i(k)$ by milling its opinion with others while the shared variable $t$ is not controllable, but only perceived and known by all the agents. Here $x_i(k)$ is updated based on the opinion dynamic process of (1). Each automaton state is updated when $t=k\Delta t$ reaches $x_i(k)$. Note that $x_i(k)$ is individually initialized. Social topology among agents plays an important role in determining how $x_i$ converges in a collective sense.

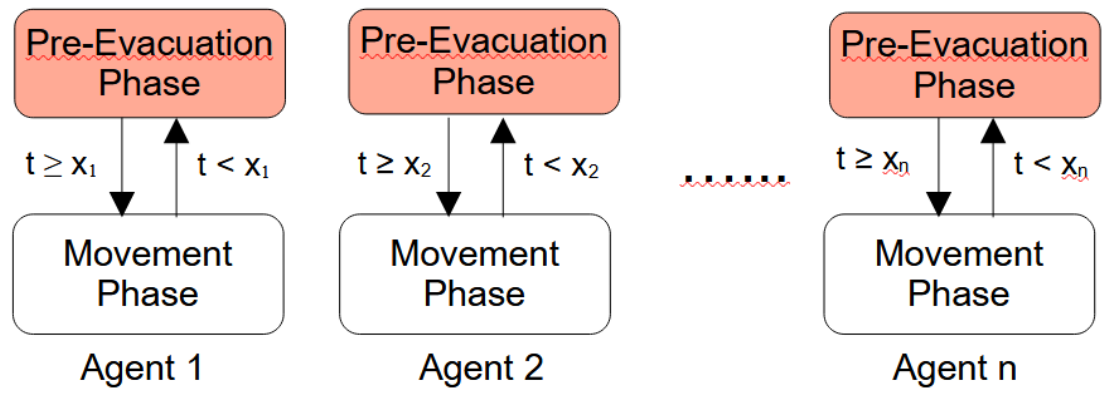

**Fig. 2. Binary Phase Transition in Interacting Extended Automata**

The above model illustrates binary phase transition of a group of interacting automata. It symbolizes the fight-or-flight response, in which individuals perceive the threat and decide in a binary manner to react to a potential or actual threat [1]. This mental process is also marked by physical adaptive changes such as nervous and endocrine changes that prepare an individual to either defend or retreat. The functions of this response were initially described in the early 1900s by American neurologist and physiologist Walter Bradford Cannon, who also coined the phrase [1]. In sum the fight-or-flight effect is the underlying cause of pre-evacuation time period for reaction to an alarm signal, explaining why evacuees often search for cues by instinct to confirm the situation instead of directly escaping for safety.

## III. Consensus and Phase Transition

If opinions of $n$ agents are vectorized by $\boldsymbol{x}(k)=[x_1(k), x_2(k), \ldots x_n(k)]^T$, we have $\boldsymbol{x}(k+1)=\mathbf{A}\cdot\boldsymbol{x}(k)$ from (1), where matrix $\mathbf{A}$ is given by $\mathbf{A}=(\boldsymbol{I}-\boldsymbol{P})+\boldsymbol{P}\cdot\boldsymbol{C}$, and $\boldsymbol{P}=diag(p_1, p_2, \ldots p_n)$ is a diagonal matrix and $\boldsymbol{I}$ is the identity matrix. Because each row of matrix $\boldsymbol{A}$ is summed up to 1, 1 is the eigenvalue of $\boldsymbol{A}$, and $\boldsymbol{A}\boldsymbol{1}_n=1\cdot\boldsymbol{1}_n$ where $\boldsymbol{1}_n=[1, 1, \ldots 1]^T$ is the corresponding rightside eigenvector with all its elements equal to 1. The system dynamics is convergent if 1 is the exclusive maximum-modulus eigenvalue of $\boldsymbol{A}$ on the unit circle, otherwise the system is periodic [9], [10], [11]. In analysis of pre-evacuation dynamics we emphasize that independent decision making is often difficult. Evacuees normally follow each other and behave in groups. Such social group behavior implies convergence of opinions such that evacuees act collectively rather than individually. Thus, the following discussion will not consider the periodic phenomenon in the French-DeGroot model. Besides, there is a well-known variation of the French-DeGroot model proposed by Friedkin and Johnsen, and it assumes that agents always keep certain prejudice in the iteration process [11], [12]. This structure emphasizes individualism in opinion dynamics, where consensus is often difficult to be reached, and thus it may not be suited to model collective response in pre-evacuation phase.

The French-DeGroot model suggests that opinions of agents converge if matrix $\boldsymbol{A}$ deduces an aperiodic graph. The system dynamics is bounded stable in case of shifting value of $0 \le c_{ij} \le 1$ and $0 \le p_i \le 1$. Changing the model parameter $c_{ij}$ , $p_i$ and the initial value $x_i(0)$ affect the final value converged and time period for convergence. Some study cases are comparatively illustrated in Fig. 1. For instance, if $c_{21}$ decreases from 0.7 to 0.1, agent 2 merges his or her opinion more towards agent 3. As a result, the final value converged is shifted upwards from roughly 12.5min to14min by comparing Fig 1(b) with Fig 1(c). The time cost for convergence is reduced. Such fast convergence does not mean that the agents respond quickly to the alarm. Instead, they fast reach an agreement that they do not want to leave for safety right away. Thus, they delay their response and their desired pre-evacuation time is increased correspondingly.

A special case of convergence is consensus, where opinions of all the agents converge to the same value as also shown in Fig. 1. Now if $p_2$ is decreased from 0.5 to 0.1, it means that agent 2 become more and more stubborn, and the converging time is shifted rightwards as illustrated in Figure 1(d). As a result, it takes longer time for the agents to reach consensus, but their desired pre-evacuation time is reduced significantly due to agent 2's stubbornness, simply because agent 2 is stubborn to a correct decision. Thus, they do not rapidly reach a consensus, but they do respond more actively and quickly to the outside.

By comparing Fig. 1(c) and Fig. 1(d), another important issue is whether the agents reach consensus before or after the crossing point with the dash line. If consensus is reached before the crossing point, it means that agents decide to escape together as a social group. If not, it may suggest that they decide to escape individually, and the consensus does not make a common sense as in the French-DeGroot Model. For example, Fig. 1(d) illustrates that people reach a consensus of leaving for safety in 10 minutes after receiving the alarm signal, but such opinions converge after 11 minutes in the simulation time horizon. Because time proceeds irreversibly, there is no way to turn the time back at 10 minutes to realize their decision in consensus. This situation actually means that people may either respond individually because they cannot reach agreement efficiently, or simply delay their decision till a consensus is reached for all of them. Thus, we extend the above phase transition model for two possible scenarios:

- **Phase Transition Type 1:** Timing is prior to consensus. This situation means that people may respond individually if they cannot reach agreement in an efficient manner. Timing is thus more important than consensus. As illustrated in Fig. 1(d), the phase transition occurs whenever the dash line goes exactly across the colorful lines no matter whether people are able to reach consensus or not before the crossing point.
- **Phase Transition Type 2:** Consensus is prior to timing. This situation means that people with strong relational ties must first reach agreement before phase transition. If the consensus is reached after the crossing point of the colorful lines and dash line, the phase transition occurs as soon as the consensus is reached. However, the content of consensus is simply meaningless because people cannot turn the time back to realize their late consensus. This type means that people delay their response till a consensus is reached.

No matter whether Type 1 or Type 2 transition is applied, we emphasize that the late consensus is partly meaningless if it happens after the crossing point of dash line and the colorful lines, and people cannot turn the time back to realize the content of consensus in the physics world. Thus, we define the effective consensus for the pre-evacuation dynamics as below.

**Definition**: Effective consensus is reached if the consensus value is no smaller than the time spent to reach the consensus.

The consensus value is computed based on the property of the French-DeGroot Model. That is, there exists a unique positive row vector $\boldsymbol{\omega}=[\omega_1, \omega_2 ... \omega_n]$ which is also the leftside dominant eigenvector of matrix $\boldsymbol{A}$ such that $\lim_{k\to+\infty}\boldsymbol{A}^k=\boldsymbol{1}_n\boldsymbol{\omega}$. Note that $\boldsymbol{\omega A}=\boldsymbol{\omega}$, and $\boldsymbol{\omega}$ is row stochastic, i.e., $\omega_1+\omega_2+...\omega_n=1$. As a result, the consensus of the French-DeGroot model exists if 1 is the simple maximum-modulus eigenvalue of $\boldsymbol{A}$, and the value of consensus is obtained by $\boldsymbol{\omega}\mathbf{x}(0)$.

The time slot required to reach a consensus is called the converging time of the above model, which is determined by the second modulus-largest eigenvalues of $\boldsymbol{A}$, whose modulus is exactly less than 1. If all the eigenvalues of $\boldsymbol{A}$ are sequenced as $\lambda_1=1, \lambda_2, \lambda_3, ... \lambda_n$, and $|\lambda_1|=1 > |\lambda_2| \ge |\lambda_3| \ge ... \ge\lambda_n$ . The converging time exactly depends on $|\lambda_2|<1$ with $\lim_{k\to+\infty}|\lambda_2|^k=0$. Given $0<\varepsilon<1$ as the accepted error bound for convergence, (e.g., $10^{-2}$), $|\lambda_2|^k<\varepsilon$ determines the converging time, which is estimated by $\Delta t\ln\varepsilon/\ln|\lambda_2|$. Therefore, the criterion of judging if a consensus is effective is given by $\boldsymbol{\omega}\mathbf{x}(0)>\Delta t\ln\varepsilon/\ln|\lambda_2|$. Here we mention that the desired pre-evacuation time for $n$ agents are initialized by $x_i(k)>0$ at $t=0$ such that $\boldsymbol{\omega}\mathbf{x}(0)>0$, and $\Delta t$ is measured by the same time unit with elements in $\mathbf{x}(0)$.

## IV. Vicsek Flocking Model and Behavioral Simulation

In this section we introduce another important feature of crowd escape - herding effect, within a general framework of the Vicsek flocking model. Herding is evident in evacuation events because dependent decision making is often difficult due to excessive time-pressure in a stressful condition. By instinct, people normally follow each other rather than act independently. Such herding and flocking behavior is known as a primary phenomenon, not only for human crowd, but also for many social animals, e.g., flock, herd and school. If we look into a real-life scenario of evacuation such as in a stadium or shopping malls, where many people follow the crowd flow to move to exits. There are a multitude of small groups composed by family members or friends, and they keep together due to their relational ties. These small groups compose a large flow of crowd, and herding behavior widely exists in both of intra-group and inter-group dynamics. The interpersonal ties (e.g., family members or friends) should be characterized by $\boldsymbol{C}=[c_{ij}]_{n\times n}$ , and a general effect of herding exists among such groups, emphasizing how people tends to follow others within a confidence range. Thus, this section will extend (1) slightly such that the social connectivity is time-varying and a binary connectivity weight $\varphi_{ij}$ is integrated with $c_{ij}$ as below.

$$x_i(k+1) = (1-p_i)x_i(k)+p_i (\Sigma_j c_{ij} \varphi_{ij}(k) x_j(k)) / (\Sigma_j c_{ij} \varphi_{ij}(k))$$
$$= x_i(k) + p_i (\Sigma_j c_{ij} \varphi_{ij}(k)[x_j(k) - x_i(k)]) / (\Sigma_j c_{ij} \varphi_{ij}(k))$$
$$\varphi_{ij}(k) = 1 \text{ if agent } i \text{ has access to agent } j\text{'s opinion}$$
$$\varphi_{ij}(k) = 0 \text{ if agent } i \text{ has no access to agent } j\text{'s opinion} \quad (2)$$

In (2) $\varphi_{ij}(k)$ is a binary variable, either 1 or 0, describing if agent $i$ has access to opinion of agent $j$ (e.g., communicating to agent j) at time step k. $\boldsymbol{\Phi}(k)=[\varphi_{ij}(k)]_{n\times n}$ is a binary matrix for connectivity of $n$ agents, determining how $\mathbf{C}=[c_{ij}]_{n\times n}$ is counted in computation at time step k. For consistency with (1), we still confine $0\le c_{ij}\le 1$ and $\Sigma_j c_{ij}=1$ throughout the paper, but this is actually not required for (2) in practical computing. Here $c_{ij}$ in (2) can be extended in domain of [0, +∞), and its value is merely a relative measurement for relational ties among individuals. The corresponding weight is normalized within [0, 1] in computational loops automatically by dividing $\Sigma_j c_{ij}\varphi_{ij}(k)$, and in practical computing we often include self-loops in the above model such as $c_{ii}=10^{-10}$ and $\varphi_{ii}(k)=1$ for ∀i=1,2,...n, ensuring that $\sum_{j=1\ldots n} c_{ij}\varphi_{ij}(k)>0$ and self-loop is only counted effectively for any isolated agents. Furthermore, from the perspective of graphs, (2) inherently defines two graphs, which are G[**C**]=(V, E[**C**]) and G[**Φ**(k)]=(V, E[**Φ**(k)]). The vertex in V={1,2,... n} is in one-to-one correspondence with each agent. According to (2), a major study topic is how $\varphi_{ij}$ is determined by combination of agent states. For instance, the bounded confidence model assumes that each agent possesses a confidence range, and $\varphi_{ij}(k)=1$ within the range [13]. The model is opinion-dependent as a nonlinear dynamics.

$$\varphi_{ij}(k)=1 \text{ if } | x_i(k)-x_j(k) |\le R; \quad \varphi_{ij}(k)=0 \text{ otherwise} \quad (3)$$

The Vicsek flocking model [13], [14], [15] is a simplified version of Boids model [16], and it assumes that a group of agents steer towards the average heading of local others who are sufficiently close to them. Their headings converge if the flock density exceeds a certain limit. In our application we are interested in convergence of desired pre-evacuation time $x_i(k)$ instead of heading vectors. In order to integrate (2) in the Vicsek flocking framework, it is necessary to integrate the multi-agent opinion process (2) with binary phase-transition rule into a more general framework of many-particle simulation in a planar space, which is widely used in existing crowd simulation [17], [18], [19]. The motion equation is given by physics laws of Langevin Equation as below.

$$m_i (d\boldsymbol{v}_i/dt) = \boldsymbol{E} - \gamma\boldsymbol{v}_i \quad (d\boldsymbol{r}_i/dt) = \boldsymbol{v}_i \quad \Rightarrow \quad \boldsymbol{v}_i(k+1)=\boldsymbol{v}_i(k) + (\boldsymbol{E} - \gamma\boldsymbol{v}_i(k))\Delta t / m_i \quad \boldsymbol{r}_i(k+1)=\boldsymbol{r}_i(k)+ \boldsymbol{v}_i(k)\Delta t \quad (4)$$

where $m_i$ is the agent mass, and $\boldsymbol{r}_i(t)$ or $\boldsymbol{v}_i(t)$ are the agent's position and velocity in a two-dimensional space at instantaneous time $t$. $\boldsymbol{E}$ is an external egress field that guides agent $i$ moving toward a target location such as an exit or a safe place. It is defined to describe crowd motion in a movement phase [17], [19]. In the pre-evacuation phase $\boldsymbol{E}$ is simply given zero. For simplicity, the Eular's method is adopted to discretize (4) as widely used in the Vicsek flocking model. The Vicsek flocking model assumes spatial proximity to determine $\varphi_{ij}(k)$, and $\varphi_{ij}(k)=1$ if the spatial distance of two agents is within an interaction range $l$,

$$\varphi_{ij}(k)=1 \text{ if } \quad d_{ij}(k)=|\mathbf{r}_i(k)-\mathbf{r}_j(k)|< l; \quad \varphi_{ij}(k)=0 \text{ otherwise} \quad (5)$$

Thus, if agent $j$ move into a circle of radius $l$ surrounding the agent $i$, the opinion of agent $i$ is influenced by agent $j$. Given a system composed by $n$ agents, and each agent state is denoted by $\{\boldsymbol{r}_i(k), \boldsymbol{v}_i(k), x_i(k)\}$. The entire system state space is the union of all the agent's state. The phase transition for each agent is determined by comparing the value of $k\Delta t$ and $x_i(k)$. The opinion dynamics $x_i(k)$ for each agent is formulated by (2), and $\boldsymbol{r}_i(k)$ and $\boldsymbol{v}_i(k)$ are updated based on physics laws of (4), which is discretized by the Eular method as in the Vicsek flocking model. The opinion dynamics (2) is coupled with many-particle system (4) through a logical rule (5). A numerical testing result of multiple evacuees is thus illustrated in Fig. 3, where a multi-compartment layout is created based on MassEgress simulation [18]. The simulation result was obtained by using FDS+Evac [18], where opinion model (2) is applied with a many-particle system (4) and criterion $d_{ij}(k)<l$ is used to enable interaction among multiple evacuee agents. The simulation result is further exhibited by 3D visualizer Smokeview [20] as below.

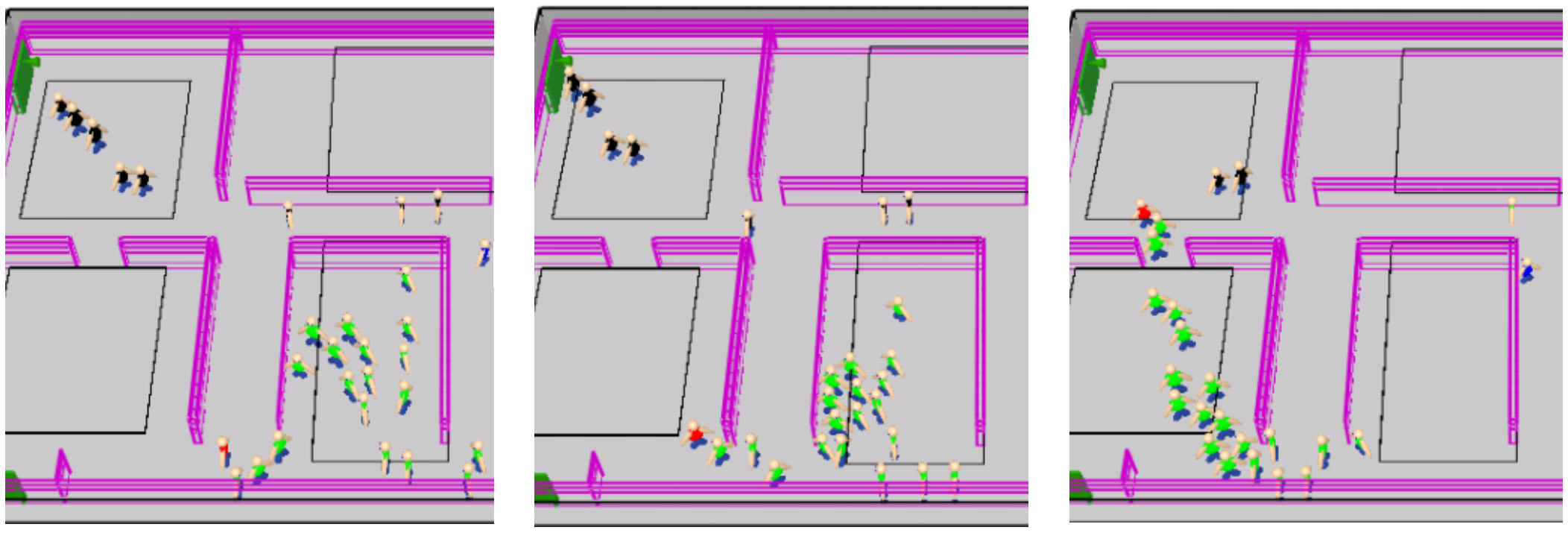

**(a)** **(b)** **(c)**

**Figure 3. Herding Effect in Pre-Evacuation Phase and Movement Phase in Multi-Compartment Layout.**

In Fig. 3 opinions of agents interact if they are spatially close enough, i.e., $d_{ij}(k)<l$. The red evacuee agent is initialized with desired pre-evacuation time $x_i(t)$ much smaller than the green ones and (1), (2) are jointly effective for their opinion exchange. As a result, the desired pre-evacuation time converges for agents in both colors. As simulation time $t$ reaches their common desired pre-evacuation time, the entire group of green agents move together with the red one to an exit as shown in Fig 3(c).

In sum herding widely exists in crowd evacuation, and the Vicsek flocking model implies this dynamical feature, and this is the underlying reason why the Vicsek flocking model is suited in this study subject. However, human crowd is much more complex in social structure than flocks and herds, and this motivate us to advance the Vicsek flocking model in several aspects. A straightforward idea is assuming cohesion effect as in Boids model [16], where evacuees are clustered by a cohesive force. The cohesive force helps to achieve physical proximity for certain individuals in close relationship, e.g., friends and family members, increasing the probability of satisfying the criterion $d_{ij}(k)<l$. As a result, people find their trust ones for opinion exchange and collective decision [2], [21], and such behavior is also evident in socio-psychological study, as called social attachment effect [21].

Last but not least, how to initialize the pre-evacuation time in the simulation is another challenging task. Existing approach posits that pre-evacuation time follow certain probability distributions, such as Lognormal, Loglogistic or Weibull distribution [22], [23]. Such statistical models are useful to explore whether or to what extent the pre-evacuation time is related to certain factors such as the location of egress, e.g., shopping malls or office buildings, and thus may provide a guideline to initialize the desired pre-evacuation time $x_i(k)$ in our modeling framework. However, a difficulty may exist in acquiring reliable data in real-world events or virtual-reality experiments [24], [25], [26]. Effectiveness of such data need further validation. Besides, we are also calling for empirical data to test our model quantitatively or calibrate the model parameters on the foundation of social science. These tasks are also identified as our future work.

## V. Conclusion

This paper introduces a binary decision process to study self-organizing phenomenon of evacuees in an evacuation process. The model is established as a mixture of physics models and dynamical systems with socio-psychological theories referring to the fight-or-flight response, social groups and herding instinct. The social topology of evacuees plays an important role in their group decision making, and the simulation result agrees with existing theory in classic opinion dynamics. The model is meaningful to bridge a gap between engineering practices and social psychological findings in egress research, and it will enable us to study a kind of complex social interaction, not only in the field of evacuation research, but also in an cross-disciplinary field of statistical physics and computational social science.

## Acknowledgments

This paper is especially dedicated to Prof. Peter B. Luh for memorizing his insightful guidance in this study subject. The authors are sincerely thankful to Prof. Meng-Chu Zhou, Prof. Kerry L. Marsh, Prof. Xiaoda Wang, Neal Olderman, Shi-Chung Chang, Bo Xiong and Vivek Kant for helpful discussion.